\documentstyle[12pt]{article}

\begin{document}

\newcommand\vect[1]{{\mbox{\boldmath $#1$}}}
\newcommand\sfrac[2]{{\textstyle{\frac{#1}{#2}}}}

\newcommand\deriv[2]{\displaystyle\frac{\partial #1}{\partial #2} }
\newcommand\col[2]{ \left(\begin{array}{c}#1\\#2\end{array}\right) }
\newcommand\oper[1]{{\cal #1}}

\newcommand{\la}{\;
  \raise0.3ex\hbox{$<$\kern-0.75em\raise-1.1ex\hbox{$\sim$
  }}\;\hskip-2pt }
\newcommand{\ga}{\;
  \raise0.3ex\hbox{$>$\kern-0.75em\raise-1.1ex\hbox{$\sim$
  }}\;\hskip-2pt }
\newcommand\bgreek[1]{{\mbox{\boldmath $#1$}}}

%
%
\newcommand{\cm}{\,{\rm cm}}
\newcommand{\cmcube}{\,{\rm cm^{-3}}}
\newcommand{\dyn}{\,{\rm dyn}}
\newcommand{\erg}{\,{\rm erg}}
\newcommand{\Jy}{\,{\rm Jy}}
\newcommand{\Jyb}{\,{\rm Jy/beam}}
\newcommand{\kms}{\,{\rm km\,s^{-1}}}
\newcommand{\mJy}{\,{\rm mJy}}
\newcommand{\mJyb}{\,{\rm mJy/beam}}
\newcommand{\K}{\,{\rm K}}
\newcommand{\kpc}{\,{\rm kpc}}
\newcommand{\Mpc}{\,{\rm Mpc}}
\newcommand{\mG}{\,{\rm mG}}
\newcommand{\mkG}{\,\mu{\rm G}}
\newcommand{\MHz}{\, {\rm MHz}}
\newcommand{\Msol}{\,{\rm M_\odot}}
\newcommand{\p}{\,{\rm pc}}
\newcommand{\radm}{\,{\rm rad\,m^{-2}}}
\newcommand{\s}{\,{\rm s}}
\newcommand{\yr}{\,{\rm yr}}

\title{\Large{\bf NON-LOCAL EFFECTS IN THE MEAN-FIELD DISC DYNAMO.\\ I.
AN ASYMPTOTIC EXPANSION}}

\author{
VLADIMIR PRIKLONSKY\\
{\normalsize\it Department of Physics, Moscow University, Moscow
119899,
Russia}\\[10pt]
ANVAR SHUKUROV\thanks{Corresponding author. e-mail: Anvar.Shukurov@Newcastle.ac.uk}\\
{\normalsize\it Department of Mathematics, University of
Newcastle,}\\
{\normalsize\it Newcastle upon Tyne NE1~7RU, UK}\\[10pt]
DMITRY SOKOLOFF\\
{\normalsize\it Department of Physics, Moscow University, Moscow
119899,
Russia}\\[10pt]
ANDREW SOWARD\\
{\normalsize\it School of Mathematical Sciences, University of
Exeter,}\\
{\normalsize\it Exeter EX4 4QE, UK}\\[10pt]
}

\date{}
\maketitle
\bigskip
\centerline{
{\normalsize\it (Received 2 August 1999; in final form 25 February 2000)}
}

\begin{abstract}We reconsider thin-disc global asymptotics for kinematic,
axisymmetric mean-field dynamos with vacuum boundary conditions.
Non-local terms arising from a small
but finite radial field component at the disc surface are consistently
taken into account for quadrupole modes. As in earlier approaches, the
solution splits into a local part describing the field distribution
along the vertical direction and a radial part describing the radial
(global) variation of the eigenfunction. However, the radial part of
the eigenfunction is now governed by an integro-differential equation
whose kernel has a weak (logarithmic) singularity. The integral term
arises from non-local interactions of magnetic fields at different
radii through vacuum outside the disc. The non-local interaction
can have a stronger effect on the solution than the local
radial diffusion in a thin disc, however the effect of the integral
term is still qualitatively similar to magnetic diffusion.

\bigskip

\end{abstract}

\section{Introduction}

In the standard thin-disc asymptotic approach to mean-field
dynamos, the dynamo equation
\begin{equation}                        \label{MFD}
\deriv{\vect{B}}{t}
=\nabla\times\left(\vect{V\times B}+\alpha\vect{B}
-\beta\nabla\times\vect{B}\right)\;,
\end{equation}
splits into a local part containing only derivatives with respect to the
vertical coordinate $z$ and the horizontal part containing
derivatives with respect to the cylindrical radius $r$ and the
azimuthal angle $\phi$. Here $\vect{B}$ is the mean magnetic field,
$\vect{V}$ is the mean velocity, $\alpha$ is the coefficient
describing the alpha-effect, and $\beta$ is the turbulent
magnetic diffusivity. We use cylindrical polar coordinates
$(r,\phi,z)$ and consider axisymmetric solutions of the kinematic
dynamo equation, $\partial/\partial{\phi}=0$. In terms of the
azimuthal components of the magnetic field $B$ and the vector
potential $A$, Eq.~(\ref{MFD}) reduces to
\begin{eqnarray}
\deriv{B}{t}&=&-DG\deriv{A}{z}+\deriv{^2B}{z^2}
+\lambda^2\deriv{}{r}\left(\frac{1}{r}\deriv{}{r}rB\right)\;,
                                                 \label{MFDB}\\
\deriv{A}{t}&=&\alpha B+\deriv{^2A}{z^2}
+\lambda^2\deriv{}{r}\left(\frac{1}{r}\deriv{}{r}rA\right)\;,
                                                 \label{MFDA}
\end{eqnarray}
where $D=\alpha_0 Gh^3/\beta^2$ is the dynamo number,
$G=r\,\partial\Omega/\partial r$ (with $\Omega$ the angular
velocity), and we have used dimensionless variables defined by the
disc semi-thickness $h$ for $z$, characteristic disc radius $R$ for
$r$, magnetic diffusion time $h^2/\beta$ for $t$, and the ratio of
units for $B$ and $A$ is $\alpha_0h^2/\beta$,
with $\alpha_0$ a representative value of the $\alpha$-coefficient (see, e.g.,
Ruzmaikin {\it et al.,} 1988 for details).  We have neglected the
term with $\partial\Omega/\partial z$ in Eq.~(\ref{MFDB}) assuming
that $|\partial\Omega/\partial z|\ll|\partial\Omega/\partial r|$
under typical conditions in astrophysical discs. In terms of the
dimensionless variables, we consider the region $|z|\leq1,\
0<r<\infty$. Astrophysical discs have no well defined radial
boundary, and the dynamo active region has a finite size because the
intensity of the dynamo action decreases at large $r$; therefore, the
characteristic radius of the dynamo region $R$, introduced below, is
finite. The disc boundary is not sharp at its surface as well, but
the turbulent magnetic diffusivity rapidly increases with height (at
least, in spiral galaxies), so that vacuum boundary conditions posed
at a finite height are a good approximation (Moss {\it et al.},
1998).

Equations (\ref{MFDB}) and (\ref{MFDA}) are written for the
$\alpha\omega$-dynamo, but generalization to $\alpha^2\omega$-dynamos
is straightforward as long as the radial scale of the solution
in not affected much. Our basic result, Eq.~(\ref{finaleq}), remains
valid for both $\alpha\omega$- and $\alpha^2\omega$-dynamos, but
$\gamma_0(r)$ and $\eta(r)$ defined below will follow from different local
equations.

An asymptotic solution of Eqs.~(\ref{MFDB}) and (\ref{MFDA}) is
represented by Ruzmaikin {\it et al.}\ (1988) in the form
\[
\col{B}{A}=\exp{(\Gamma t)}\left[Q(\lambda^{-1/2}r)\col{b(z;r)}{a(z;r)}+\ldots\right]\;,
\]
where $\Gamma$ is the growth rate of the field ($\Gamma$ is known to
be real for the dominant modes in a thin disc if $D<0$), $(b,a)$ is the
local solution (suitably normalized) depending on $r$ only
parametrically, and $Q(r)$ is the amplitude of the global solution.
The asymptotic parameter is the disc aspect ratio
\[
\lambda=\frac{h}{R}\simeq10^{-2}\mbox{--}10^{-1}\;.
\]

Equation (\ref{MFD}) should be supplemented by appropriate boundary
conditions (discussed in detail below), and then the lowest-order
approximation in $\lambda$ yields a boundary value problem for
$(b,a)$, whereas the solvability condition for the first-order
equations leads to an equation for $Q$ and $\Gamma$. The procedure for
the derivation of the asymptotic equations is reconsidered in detail
below.

The boundary conditions traditionally used in conjunction with
Eqs.~(\ref{MFDB}) and (\ref{MFDA}) are the so-called vacuum boundary
conditions. If there are no electric currents outside the disc, then
$\nabla\times\vect{B}=0$ and, for axisymmetric solutions,
(see, e.g., Zeldovich {\it et al.}, 1983, p.\ 151)
\[
B=0\quad\mbox{for}\quad |z|\geq1\;.
\]
Further, $\nabla^2\vect{A}=0$ outside the disc, so that
\begin{equation}
\deriv{^2A}{z^2}+\lambda^2
\deriv{}{r}\left(\frac{1}{r}\deriv{}{r}rA\right)=0
\quad\mbox{for}\quad |z|\geq1\;.   \label{BA}
\end{equation}
In
order to impose boundary conditions at the disc surface, we recall
that the eigenfunctions of Eq.~(\ref{BA}) are of the form
$A\propto\exp{(-k\lambda|z|)}J_1(kr)$ with some radial wave number $k$.
Since we require that the scale of the radial variation of $A$ is of
order unity, $k=O(1)$, we obtain
\begin{equation}
\deriv{A}{z}=-k\lambda A=O(\lambda)\quad\mbox{for}\quad
z\geq1\;.\label{BdA}
\end{equation}
To the lowest order in $\lambda$, the resulting boundary condition is
\begin{equation}
\deriv{A}{z}=0\quad\mbox{for}\quad z=1\;.\label{BdA0}
\end{equation}

It was noted by Soward (1992a,b) that the term of order $\lambda$ in
Eq.~(\ref{BdA}) must be retained if higher-order terms of the
asymptotic solution of Eqs.~(\ref{MFDB}) and (\ref{MFDA}) are
considered.  Our goal here is to include these terms in a consistent
manner into a regular asymptotic scheme. The physical meaning of
these corrections is as follows.  Since $-\partial A/\partial z=B_r$,
the boundary condition (\ref{BdA0}), together with $B=0$, restricts
the  magnetic field to be purely vertical outside the disc, similarly
to an external field of an infinite homogeneous slab.  Then magnetic
fields at different radii can be connected only via magnetic lines
passing through the disc where $B_r\neq0$.  However, the boundary
condition (\ref{BdA}) implies that the field on and above the disc
surface has a weak but finite radial component, and magnetic
lines can thus connect different regions in the disc through the
vacuum. This leads to non-local effects arising in higher orders of
asymptotic expansion, and a description of these effects is our subject
here.

A treatment of the non-local asymptotics for mean-field
dynamos in a thin disc can be found in Soward (1978, 1992a,b, 2000)
where both steady and oscillatory dipolar and quadrupolar modes
are considered, but only for marginally stable solutions and in a
local, Cartesian geometry. Here we consider exponentially growing
(or decaying), non-oscillatory, quadrupolar modes in cylindrical
geometry, ready for applications to astrophysical discs, especially
those in spiral galaxies. Another new element in our approach is that
we do not expand the coefficients of the dynamo equations into power
series in $r$, and so we allow for arbitrary radial variation in the
dynamo coefficients. In this paper we present the derivation of the
asymptotic equations. Their solutions will be discussed elsewhere
(Willis {\it et al.,} 2000).

\section{Vacuum boundary conditions for a thin disc} \label{VBC}

In this section we derive boundary conditions for a thin disc
surrounded by vacuum in a form suitable for asymptotic analysis.
For this purpose we first perform the Hankel transform of
Eq.~(\ref{BA}) to obtain:
\begin{equation}
\deriv{^2\widehat A}{z^2}-\lambda^2 k^2 \widehat A=0\;,\label{BhatA}
\end{equation}
where
\begin{equation}
\widehat A(z,k)=\int_0^\infty A(z,r)J_1(kr)r\,dr \label{hatA}
\end{equation}
is the Hankel transform of $A(z,r)$ with respect to $r$.
Equation (\ref{BhatA}) has a simple solution decaying at infinity,
$\widehat A(z,k)=A_0(k)\exp{(-k\lambda|z|)}$.  However, this solution
contains an arbitrary function of $k$, $A_0(k)$, so it cannot be used
to formulate a closed boundary condition at $|z|=1$.  Instead, we
note that, for $z>0$, solution of Eq.~(\ref{BhatA}) decaying at
$z\to\infty$ satisfies
\[
\deriv{\widehat A}{z}+\lambda k\widehat A=0\quad\mbox{for }z\geq1\;,
\]
which is useful to compare with Eq.~(\ref{BdA}). We apply
the inverse Hankel transform to this equation and set $z=1$ to obtain
\begin{equation}
\left.\deriv{A}{z}\right|_{(1,\,r)}+\lambda\int_0^\infty
\widehat A(1,k)J_1(kr)k^2\,dk=0\;,      \label{BAn}
\end{equation}
and the plus sign must be replaced by minus for $z=-1$.

We could now substitute Eq.~(\ref{hatA}) into Eq.~(\ref{BAn}) in
order to obtain a closed boundary condition for $A$ at $z=1$ in an
integro-differential form, but the resulting operator contains a
strongly divergent kernel $\int_0^\infty J_1(kr)J_1(kr')k^2\,dk$.
Therefore, we develop a slightly more elaborate procedure devised to
isolate the singularity in the integral kernel. Firstly, we improve the
convergence of the kernel by eliminating the factor $k^2$ in it.
This can be done by rewriting the integral operator in terms of
$\partial{^2A}/{\partial r^2}$. For this purpose, we use the general
relation (Sneddon, 1951, Sect.~10)
\[
k^2\widehat A = -\int_0^\infty
\deriv{}{r}\left( \frac{1}{r}\,\deriv{}{r}rA\right) J_1(kr)r\,dr\;,
\]
written at $|z|=1$, to reduce Eq.~(\ref{BAn}) to the form
\begin{equation}
\deriv{A}{z}-\frac{\lambda}{r}\int_0^\infty
      \deriv{}{r'}\left(\frac{1}{r'}\,\deriv{}{r'}r'A\right)
                W(r,r')\,dr'=0\quad \mbox{at }z=1\;,
                                \label{FBA}
\end{equation}
where
\begin{equation}
W(r,r')=rr'\int_0^\infty J_1(kr)J_1(kr')\,dk\;;
                                \label{kernel}
\end{equation}
note that we have isolated a factor $rr'$ in $W(r,r')$
to ensure that it does not diverge at $r,r'\to0$ (see below).

We can prove that $W(r,r')$ is square integrable and
has only a weak singularity at $r-r'=0$. For this purpose we
explicitly isolate the singularity of $W(r,r')$ which occurs at
$r-r'=0$. We shall consider consecutively the cases prone to a
singular behaviour.

We start with $r,r'\neq0$ when a singularity at $r=r'$ can be
expected. We introduce a new variable $\sigma=kr$ for $r'>r$ or
$\sigma=kr'$ for $r'<r$ in Eq.~(\ref{kernel})
and use Eq.~(\ref{Jsing}) of Appendix~A to see that
\begin{equation}
W(r,r')=-\frac{1}{\pi}(rr')^{1/2}\ln|r-r'|+\ldots\;;
                                \label{Wsing}
\end{equation}
here and henceforth dots stand for a nonsingular part. In a similar
way, the case when $r,r'\to0$ can be shown to lead to the singularity
of Eq.~(\ref{Wsing}) because terms like $(rr')^{1/2}\ln r$ are not
singular.

Next consider $r=\mbox{const}\neq0, \, r'\to0$. The
new variable is now $\sigma=kr'$, so that
\[
W(r,r')=-\frac{1}{\pi}(rr')^{1/2}\ln|r-r'| +
\frac{1}{\pi}(rr')^{1/2}\ln r'+\ldots\;.
\]
Both logarithmic terms are finite due to the factor
$rr'$ in $W(r,r')$, so in this case the kernel is free of singularities.
As $r$ and $r'$ enter $W$ in a symmetrical manner, we conclude that
there is no singularity at $r\to0,\ r'=\mbox{const}\neq0$, as well.

Altogether, Eq.~(\ref{Wsing}) shows that the integral kernel has an
integrable (logarithmic) singularity, and thus
it can be treated as a finite kernel (see, e.g., Kolmogorov and
Fomin, 1957).  As a result Eq.~(\ref{FBA}) provides a viable exact
boundary condition for a slab surrounded by vacuum.

\section{Representation of the Asymptotic Solution}

In this section we consider asymptotic solutions of Eqs.~(\ref{MFDB})
and (\ref{MFDA}) which are free of  boundary layers.  Such layers may
arise at the disc surface and also at certain radii (e.g.,  contrast
structures --- Belyanin {\it et al.,} 1994).

With the definition
$\vect{Y}=\left(\begin{array}{c}B(z,r)\\A(z,r)\end{array}\right)$,
the eigenvalue problem for Eqs.~(\ref{MFDB}) and (\ref{MFDA}) can be
conveniently rewritten in the form
\begin{equation}                        \label{eq5}
\Gamma\vect{Y}=\oper{L}_z\vect{Y}+\lambda^2\oper{L}_r\vect{Y}\;,
\end{equation}
with the boundary conditions
\begin{equation}                \label{eq6}
  \oper{P}_{z,0}\vect{Y}= 0\quad\mbox{at }z=0\;; \qquad
\oper{P}_{z,1}\vect{Y}-\frac{\lambda}{r}
                \oper{K}\vect{Y}=0\quad\mbox{at }z=1\;,
\end{equation}
where the following operators have been introduced:
\[
\oper{L}_z =
\left(
  \begin{array}{cc}
  \partial^2/\partial{z^2} & -DG\partial/\partial{z} \\
  \alpha                   &\partial^2/\partial{z^2}
  \end{array}
   \right),\qquad
\oper L_r{X} =
\deriv{}{r}\left({1\over r}\deriv{}{r}r{X}\right),
\]
\[
\oper P_{z,0} =
\left( \begin{array}{cc}
\partial/\partial z & 0 \\
  0 & 1
  \end{array} \right), \qquad
\oper P_{z,1} =
 \left( \begin{array}{cc}
 1 & 0 \\
 0 & \partial/\partial z
\end{array}\right),
\]
and
\begin{equation}                                \label{eq7}
\oper{K}\col{x_1}{x_2}=
\col{0}{\displaystyle\int_0^\infty
dr'\,W(r,r')\oper{L}_{r'}(x_2)}\;.
\end{equation}
The operators $\oper P_{z,1}$ and $\oper{ K}$
are defined at $z=1$ and $\oper P_{z,0}$ at $z=0$.

Consider asymptotic solutions of Eqs.~(\ref{MFDB}) and
(\ref{MFDA}) for $\lambda\ll1$. We denote the asymptotic parameter
$\epsilon$, a function of $\lambda$ to be determined below, and
represent the asymptotic solution in the form
\begin{equation}
\vect{Y}(z,r)=Q(\epsilon^{-1}r) \vect{Y}_0(z;r)+\epsilon^2
\vect{Y}_1(z,\epsilon^{-1}r;r)+\ldots\;,        \label{AsEx}
\end{equation}
with unknown functions
\[
Q(\epsilon^{-1}r)\;,\quad
\vect{Y}_0=\col{b(z;r)}{a(z;r)},\quad
\vect{Y}_1=\col{B_1(z,\epsilon^{-1}r;r)}{A_1(z,\epsilon^{-1}r;r)}.
\]
Our motivation for the particular choice of this form will
immediately become clear. We only note here that Eq.~(\ref{AsEx})
features two types of dependence on $r$. One of them is connected
with the presence of the radial derivatives in Eq.~(\ref{eq5}), this
involves the argument $\epsilon^{-1}r$. Another kind of
$r$-dependence arises from the radial variation of the coefficients
of the local operator, $\oper L_z$; this is a slow parametric
variation of the solution with $r$ unrelated to any differential
operator in $r$. The asymptotic expansion of the eigenvalue $\Gamma$
will emerge later.

In order to preserve a similarity to earlier asymptotics developed
for galactic dynamos, we do not introduce scaled variables, e.g.,
$\epsilon^{-1}r$. Therefore, some of our asymptotic equations, e.g.,
Eq.~(\ref{finaleq}), will formally contain terms with $\epsilon$ or
$\lambda$, but all terms in such equations will always be of the same
order of magnitude in $\lambda$.

\subsection{Zeroth-order equations}

Substituting Eq.~(\ref{AsEx}) into Eqs.~(\ref{eq5}) and (\ref{eq6})
and combining terms of equal orders in $\epsilon$ we obtain in the
leading order in $\lambda$
\begin{equation}        \label{Ezero}
\left(\Gamma\vect{Y}_0-\oper{L}_z \vect{Y}_0\right)Q=0\;,
\end{equation}
\begin{equation}                \label{Bzero}
Q\oper{P}_{z,0}\vect{Y}_0=0\;,\quad
Q\oper{P}_{z,1}\vect{Y}_0=0\;,
\end{equation}
where we have deliberately retained the factor $Q$, a function of
$\epsilon^{-1}r$. If $Q$ were cancelled, Eq.~(\ref{Ezero}) would
become intrinsically contradictory as $\Gamma$ must be independent of
$r$ whereas the coefficients of $\oper{L}_z$ depend on $r$.
To overcome this difficulty, we consider separately two
radial ranges.  At those $r$ where $Q$ is small, Eqs.~(\ref{Ezero})
and (\ref{Bzero}) are satisfied, to the required accuracy of
$O(\epsilon^2)$, just because $Q$ is small, of order $\epsilon^2$ or
less.  At radii where $Q=O(1)$ we replace Eqs.~(\ref{Ezero}) and
(\ref{Bzero}) with the following equations:
\begin{equation}
\label{Eqzero}
\gamma_0(r)\vect{Y}_0-\oper{L}_z \vect{Y}_0=0\;,
\end{equation}
\begin{equation}                \label{BCzero}
\oper{P}_{z,0}\vect{Y}_0=0\;,\quad
\oper{P}_{z,1}\vect{Y}_0=0\;,
\end{equation}
where $\gamma_0(r)$ is the local growth rate (more precisely, its
zeroth-order term), a function of radius. With
this definition of $\gamma_0(r)$, we have
\begin{equation}
\Gamma-\gamma_0(r)=O(\epsilon^2)\quad \mbox{if }
                                Q(\epsilon^{-1}r)=O(1)\;.
\label{eig}
\end{equation}
Outside the radial range where $Q(\epsilon^{-1}r)=O(1)$,
$\gamma_0(r)$ still can be defined as the leading eigenvalue of
Eqs.~(\ref{Eqzero}) and (\ref{BCzero}).

Equation (\ref{eig}) in fact restricts the radial range $\Delta
r$ where $Q=O(1)$ to be $\Delta r=O(\epsilon)$ provided $\gamma_0(r)$
can be   approximated by a quadratic function of $r$
near its maximum, $\gamma_0(r+\Delta r)\simeq \gamma_0(r)+C\Delta
r^2$ with a certain constant $C$. The radial variation of
$\gamma_0(r)$ is controlled by that of $\alpha$ and $G$, the
coefficients of $\oper{L}_z$. The weaker is the radial variation of
$\alpha$ and $G$, the wider is the radial range where Eq.~(\ref{eig})
can be satisfied.

Earlier experience with solutions of Eqs.~(\ref{Eqzero}) and
(\ref{BCzero}) (Ruzmaikin {\it et al.,} 1988) shows that the leading
local eigenvalue is real for $D<0$ and, furthermore, this eigenvalue
is isolated, i.e., there are no other positive eigenvalues for
moderate values of $|D|$ (in other words, the difference of the
eigenvalues is of order unity). We restrict our attention to the mode
with the largest real eigenvalue $\gamma_0(r)$.

Since the boundary value problem (\ref{Eqzero}) and (\ref{BCzero}) has
been extensively studied earlier (e.g., Ruzmaikin {\it et al.,} 1988 and
references therein), we do not discuss it further, but assume that
its solution, $\vect{Y}_0(z;r)$ and $\gamma_0(r)$, is given.

\subsection{First-order equations}

Having isolated zeroth-order terms, we can represent the remaining
terms in Eq.~(\ref{eq5}) as
\begin{equation}
\Gamma\vect{Y_1}=\oper{L}_z\vect{Y}_1
-\frac{\Gamma-\gamma_0(r)}{\epsilon^2}\vect{Y}_0Q
+\frac{\lambda^2}{\epsilon^2}\vect{Y}_0\oper{L}_rQ+\ldots\;,
                                        \label{first}
\end{equation}
where dots denote terms of order $\epsilon^2$ and higher. We
temporarily retain the term of the order $\lambda^2/\epsilon^2$ as
the connection of $\epsilon$ and $\lambda$ still has not been
determined; we shall show that this term must be omitted in the
first-order equations.

Similarly the terms remaining in the boundary conditions are
\begin{eqnarray*}
\oper{P}_{z,0}\vect{Y}_1+\ldots&=&0\;,\\
\oper{P}_{z,1}\vect{Y}_1&=&
\frac{\lambda}{\epsilon^2r}\oper{{ K}}(Q\vect{Y}_0) +\ldots\;.
\end{eqnarray*}
We determine $\epsilon$ from the requirement that the two terms in
the second boundary condition are of equal orders of magnitude in
$\lambda$. As shown in Eq,.~(\ref{AsEx}), the eigenfunction is
expected to be of order unity in a region whose radial extent $\Delta
r$ is of order $\epsilon$, so that $dQ/dr\simeq Q/\Delta r\simeq
Q/\epsilon$.  This is true for the leading eigenfunctions $Q$ that do not rapidly
oscillate with $r$. Since $\oper{L}_r A=O(\Delta
r^{-2})=O(\epsilon^{-2})$,  we obtain $\oper{K}(Q\vect{Y}_0)\simeq
\epsilon^{-2}\Delta r\simeq \epsilon^{-1}$ (see Appendix~B).  Thus,
we require that
\[
\frac{\lambda}{\epsilon^2}\epsilon^{-1}=1\;,
\]
or
\[
\epsilon = \lambda^{1/3}\;.
\]
Then Eq.~(\ref{eig}) implies that
\[
\Gamma-\left[\gamma_0(r)]\right|_{\rm max} = O(\lambda^{2/3})\;.
\]

Now we can see that $\lambda/\epsilon=o(1)$, so that
$\lambda^2\epsilon^{-2}\oper{L}_r(Q)=O(\lambda^{2/3})$  and the desired
first-order boundary
value problem based on Eq.~(\ref{first}) reduces to
\begin{equation}
\Gamma\vect{Y_1}=\oper{L}_z\vect{Y}_1
-\frac{\Gamma-\gamma_0(r)}{\lambda^{2/3}}\vect{Y}_0Q\;,
                                        \label{Eforder}
\end{equation}
\begin{equation}                  \label{BCforder}
\oper{P}_{z,0}\vect{Y}_1=0\;,\quad
\oper{P}_{z,1}\vect{Y}_1=
{{\lambda^{1/3}}\over{r}}\oper{{ K}}(Q\vect{Y}_0)\;.
\end{equation}

However, this form of the first-order boundary value
problem is inconvenient for further analysis. We rewrite it
to obtain a boundary value problem with homogeneous boundary
conditions.  Let us represent the solution of
Eqs.~(\ref{Eforder}) and (\ref{BCforder}) in the form
\[
\vect{Y_1}=\vect{X}_1+\vect{X}_2\;,
\]
where $\vect{X}_1$ solves the following boundary value problem:
\[
\oper{L}_z \vect{X}_1 =0\;,
\]
\[
\oper{P}_{z,0}\vect{X}_1=0\;, \quad \oper{P}_{z,1} \vect{X}_1 =
\frac{\lambda^{1/3}}{r}\oper{{K}}(Q\vect{Y}_0)\;.
\]
Then $\vect{X}_2$ is the solution of the following
boundary value problem
\begin{equation}
 \Gamma \vect{X}_2-\oper{L}_z \vect{X}_2 =
\frac{\gamma_0(r)-\Gamma}{\lambda^{2/3}}\vect{Y}_0Q - \Gamma
\vect{X}_1\;, \label{Emford}
\end{equation}
\begin{equation}
\oper{P}_{z,0}\vect{X}_2=0\;, \quad \oper{P}_{z,1} \vect{X}_2 =0\;.
                                        \label{Bmford}
\end{equation}

Here $\vect{X}_1$ is in fact a first-order correction to the
eigenfunction $\vect{Y}_0$ of Eqs.~(\ref{Ezero}) and (\ref{Bzero}) arising
from the term with $\lambda^{1/3}$ in the boundary condition (\ref{BCforder})
at $z=1$. Note that $|\vect{X}_1|=O(Q)$. The boundary value problem for
$\vect{X}_1$ can be solved straightforwardly upon solving Eq.~(\ref{finaleq})
for $q(r)=Q(r)a'(1,r)$.

It is useful to represent $\vect{Y}_1$ as a
sum of the two terms, $\vect{X}_1$ and $\vect{X}_2$, because this
results in a universal form of the left-hand sides in the local
boundary value problems in all asymptotic orders. An
alternative approach, without this representation, is discussed by
Soward (2000).

We expect that $\vect{X}_2$ has the same functional form as the
zeroth-order term in the asymptotic expansion (\ref{AsEx}), so we
take $\vect{X}_2(z;r)=Q_1(\lambda^{-1/3}r)\vect{X}_{2,0}(z;r)$. Then
Eqs.~(\ref{Emford}) and (\ref{Bmford}) reduce to a form similar to
Eqs.~(\ref{Ezero}) and (\ref{Bzero}):
\begin{equation}
\Gamma \vect{X}_{2,0}-(\oper{L}_z \vect{X}_{2,0})Q_1 =
\frac{\gamma_0(r)-\Gamma}{\lambda^{2/3}}\vect{Y}_0Q-\Gamma
\vect{X}_1\;, \label{Emmf}
\end{equation}
\begin{equation}
Q_1\oper{P}_{z,0}\vect{X}_{2,0}=0\;, \quad
Q_1\oper{P}_{z,1} \vect{X}_{2,0} =0\;,  \label{Bmmf}
\end{equation}
where, as above, we retain the factor $Q_1$ in the boundary
conditions and consider again two radial ranges.  At those radii where
$\Gamma - \gamma_0(r) = O(\lambda^{2/3})$ and $Q=O(1)$, we expect
that $Q_1 = O(1)$. In this radial range we obtain the following
first-order equations:
\begin{equation}
\Gamma \vect{X}_{2,0}-\oper{L}_z \vect{X}_{2,0} =
\left[\frac{\gamma_0(r)-\Gamma}{\lambda^{2/3}}\vect{Y}_0Q -
\Gamma \vect{X}_1\right]Q_1^{-1}\;,
\label{Ezf}
\end{equation}
\begin{equation}
\oper{P}_{z,0}\vect{X}_{2,0}=0\;,\quad\oper{P}_{z,1}\vect{X}_{2,0}
=0\;.                           \label{Bzf}
\end{equation}
These equations have the same operators on the left-hand sides as
Eqs.~(\ref{Eqzero}) and (\ref{BCzero}).  As for the radial range
where $Q_1 = o(1)$, the form of $Q_1$ is unimportant there and we do
not discuss it further.  Equations~(\ref{Ezf}) and (\ref{Bzf})
represent the desired form of the first-order boundary value problem
which we use below to obtain an equation for $Q$.

\section{The radial dynamo equation}

In this section we derive a closed equation for the radial part of
the zeroth-order eigenfunction, $Q$, which follows from the
solvability condition for the first-order boundary value problem
(\ref{Ezf}) and (\ref{Bzf}). By Fredholm's solvability
condition, the r.h.s.\ of Eq.~(\ref{Ezf}) must be orthogonal to
the eigenvector $\vect{X}^{\dag}$ of the corresponding adjoint
homogeneous problem (this problem is formulated and discussed in
Appendix~C), i.e.,
\[
\left\langle\left[\frac{\Gamma-\gamma_0(r)}{\lambda^{2/3}}\vect{Y}_0Q +
\Gamma \vect{X}_1\right]Q_1^{-1}, \vect{X}^{\dag}\right\rangle=0\;,
\]
where $\langle\vect{F},\vect{G}\rangle=\int_0^1\vect{F\cdot G}\,dz$
denotes the scalar product in the space of continuous vector
functions on the interval $0\le z\le1$. This yields the following
equation for $Q$ and $\Gamma$:
\begin{equation}
\frac{\Gamma - \gamma_0(r)}{\lambda^{2/3}}Q
\langle\vect{Y}_0, \vect{X}^{\dag}\rangle  + \Gamma
\langle\vect{X}_1, \vect{X}^{\dag}\rangle =0\;.
\label{Eqr}
\end{equation}
We recall that $\Gamma-\gamma_0(r)=O(\lambda^{2/3})$ and
$|\vect{X}_1|=O(1)$ in the radial range where $Q=O(1)$ and
$|\vect{X}_1|=o(1)$ outside this range. Therefore, to the accuracy
$O(\lambda^{2/3})$ we can replace $\Gamma$ by $\gamma_0(r)$ to obtain
\begin{eqnarray*}
\Gamma\langle\vect{X}_1,\vect{X}^{\dag}\rangle&=
        &\langle\vect{X}_1,\gamma_0(r)\vect{X}^{\dag}\rangle\\
   &=&\langle\vect{X}_1,\oper{L}_z^{\dag}\vect{X}^{\dag}\rangle \\
   &=&-a^{\dag}(1,r){\lambda^{1/3}\over{r}}\oper{K}(aQ)\;,
\end{eqnarray*}
where, in order to obtain the last equality, we performed two
integrations by parts and took into account the boundary conditions
(\ref{BCzero}) and (\ref{BCadj}).

Now Eq.~(\ref{Eqr}) can be rewritten as
\[
\Gamma Q=\gamma_0(r)Q +
\frac{\lambda a^{\dag}(1;r)}{r\langle\vect{Y}_0,\vect{X}^{\dag}\rangle}
\int_0^\infty dr'\,W(r,r') \oper{L}_{r'}
\Bigl(Qa(1,r')\Bigr)\;.  \label{Erf}
\]
This is the desired equation for magnetic field
amplitude $Q$ and global growth rate $\Gamma$.  We stress that all
terms in this equation are of the same order in $\lambda$
because $dQ/dr=O(\lambda^{-1/3})$.

To reduce this equation to a final form, we introduce
\[
q(r)=Q(r)a'(1,r)\;,
\]
to obtain
\begin{equation}
\Gamma q=\gamma_0(r)q +
\frac{\lambda}{r}\eta(r)
        \int_0^\infty dr'\,W(r,r') \oper{L}_{r'}(q)\;,
                                        \label{finaleq}
\end{equation}
where
\[
\eta(r)=
\frac{a(1,r)a^{\dag}(1,r)}{\langle\vect{Y}_0,\vect{X}^{\dag}\rangle}
\]
is a function of radius which is obtained from the zeroth-order
equations,
\[
\vect{Y}_0=\col{b(z;r)}{a(z;r)}\;,\quad
\vect{X}=\col{b^{\dag}(z;r)}{a^{\dag}(z;r)}\;.
\]
It is important to note that both $q(r)$ and $\eta(r)$ are
independent of the normalization of the local solutions.

\section{Discussion}
It is
useful to compare the main result of this paper, Eq.~(\ref{finaleq}),
with the corresponding equation resulting from using the boundary
condition (\ref{BdA0}) in all orders (see, e.g., Ruzmaikin {\it et
al.,} 1988):
\begin{equation}
\Gamma Q=\gamma_0(r)Q + \lambda^2 \oper{L}_{r}Q\;.      \label{Old}
\end{equation}
The last term on the right-hand side describes magnetic diffusion in
the radial direction.  The new feature of the theory discussed here
is that this term is replaced by an integro-differential operator.
Numerical analysis of Eq.~(\ref{finaleq}) presented elsewhere (Willis
et al., 2000) shows that the integral term results in enhanced
magnetic diffusion, so that Eqs.~(\ref{finaleq}) and (\ref{Old}) have
similar physical significance. This is understandable since the
kernel $W(r,r')$ has a singularity. If this singularity were extreme,
say, $W(r,r')\simeq\delta(r-r')$, the functional form of
Eq.~(\ref{Old}) would be recovered precisely, albeit with additional
factor $\eta(r)$ which can hardly be of major importance. However,
the actual kernel has a wider radial profile (and weaker
singularity), so the magnitude of this term is enhanced by non-local
coupling, but its nature still remains basically diffusive.

The integral, non-local character of magnetic diffusion affects the
nature of the asymptotic solution for $Q(r)$. From Eq.~(\ref{Old}), the
radial width of the eigenfunction is of the order of
$\epsilon=\lambda^{1/2}$ and $\Gamma-\gamma_0=O(\lambda)$ (Ruzmaikin
{\it et al.,} 1988), whereas Eq.~(\ref{finaleq}) corresponds to
$\epsilon=\lambda^{1/3}$ and $\Gamma-\gamma_0=O(\lambda^{2/3})$. We
should note, however, that the difference is hardly important for most
applications to astrophysical discs where, for $\lambda\simeq10^{-2}$,
the difference amounts to a factor of $\lambda^{1/6}\approx1/2$. A
wider radial distribution of the magnetic field resulting from
Eq.~(\ref{finaleq}) (Soward, 2000; Willis {\it et al.,} 2000) can
help in accelerating the amplification of magnetic fields in the
outer regions of young galaxies along the lines discussed by Moss
{\it et al.} (1998).

Using heuristic arguments, Poezd {\it et al.} (1993) generalized
Eq.~(\ref{Old}) to nonlinear regimes arising from alpha-quenching. The
resulting nonlinear equation has $\Gamma$ replaced by
$\partial/\partial t$ and $\gamma_0(r)$ by
$\gamma_0(r)[1-Q^2/B_0^2(r)]$, where $B_0(r)$ defines the dynamo
saturation level appearing in the alpha-quenching model. It is
plausible that a similar generalization is possible with
Eq.~(\ref{finaleq}). Another problem is to generalize the theory to
nonaxisymmetric solutions where the boundary conditions should take a
more complicated form.

\section*{Acknowledgements}

 The research was initiated during a one month visit (28 August to 25
 September 1993; supported by the Royal Society) of Andrew Soward to
 the Science Research Computer Center, Moscow State University.
 We are grateful to A.~D.~Poezd for his contributions at that early
stage. We thank the University of Newcastle upon Tyne for encouraging
 further mutual visits during the course of this study, which were
 supported by PPARC (grant PPA/G/S/1997/00284) and the Royal Society.

\section*{Appendix A. The singularity of $\displaystyle\int_0^\infty
J_1(k)J_1(k\sigma)\,dk$}


\setcounter{equation}{0}
\renewcommand\theequation{A\arabic{equation}}

In this section we isolate the singularity of the integral

\[
I(\sigma)=\int_0^\infty J_1(k)J_1(k\sigma)\,dk\;,
\]
where $\sigma$ is a constant. It is a special case of the Weber and Schafheitlin
integral discussed at length by Watson (1922, Sect.\ 13.4), which can expressed as
\[
I(\sigma)=\sfrac12\sigma\,{}_2F_1(\sfrac32,\sfrac12,2,\sigma^2)\;.
\]

We can restrict ourselves to the case $\sigma>1$ because
Eq.~(\ref{kernel}) reduces to the above form with $\sigma=r/r'$
for $r>r'$ and $\sigma=r'/r$ for $r'>r$.

Asymptotics of this integral for $\sigma-1\ll1$ can be obtained by
expressing $I(\sigma)$ in terms of the complete elliptic integral of the
first kind, which yields $I(\sigma)\simeq-\pi^{-1}\ln|1-\sigma|$ (see Willis {\it et
al.,} 2000 for details).

This asymptotic form can also be obtained from the following arguments which make
clear the nature of the singularity at $\sigma\to1$. Introduce
\[
\Delta(x)=J_1(x)-\left(\frac{2}{\pi x}\right)^{1/2}
        \cos\left(x-\sfrac34\pi\right)\;,
\]
so that $\Delta(x)=O(x^{-1})$ for $x\gg1$. Then
\begin{eqnarray*}
I(\sigma)&=&\int_0^1 J_1(k)J_1(k\sigma)\,dk +
        \int_1^\infty \Delta(k)\Delta(k\sigma)\,dk\\
&&\mbox{}+\int_1^\infty \Delta(k)\cos(k\sigma-\sfrac{3}{4}\pi)
                \left(\frac{2}{\pi k\sigma}\right)^{1/2}
 +\int_1^\infty \Delta(k\sigma)\cos(k-\sfrac34\pi)
                  \left(\frac{2}{\pi k}\right)^{1/2}dk\\
&&\mbox{}+2\int_1^\infty\cos(k-\sfrac34\pi)\cos(k\sigma-\sfrac34\pi)
                     \frac{dk}{\pi k\sigma^{1/2}}\;,
\end{eqnarray*}
where the only divergent term is the last one. For $\sigma>1$, the
latter integral can be expressed in terms of the sine integral  ${\rm
Si}(x)=-\int_0^x u^{-1}\sin u\,du$ and the cosine integral
${\rm Ci}(x)=-\int_x^\infty u^{-1}\cos u\,du$ (which has a singular
component $\ln x$) as
\begin{eqnarray*}
2\int_1^\infty\cos(k-\sfrac34\pi)\cos(k\sigma-\sfrac34\pi)
                                \frac{dk}{\pi k\sigma^{1/2}}
&=&\frac{1}{\pi \sigma^{1/2}}\left[-\sfrac12\pi-{\rm Si}(\sigma+1)
-{\rm Ci}(\sigma-1)\right]\\
&=&f(\sigma)-\frac{1}{\pi \sigma^{1/2}}\ln{(\sigma-1)}\;,
\end{eqnarray*}
where $f(\sigma)$ is a finite function.

Hence, for $\sigma>1$ we have
\begin{equation}
\int_0^\infty J_1(k)J_1(k\sigma)\,dk=F(\sigma) -\frac{1}{\pi
\sigma^{1/2}}\ln{(\sigma-1)}\;,          \label{Jsing}
\end{equation}
where $F(\sigma)$ is finite.

\section*{Appendix B. The order of magnitude of the integral term}

\setcounter{equation}{0}
\renewcommand\theequation{B\arabic{equation}}

Here we estimate the order of magnitude of
\[
\oper{K}(Qa) =\int_0^\infty
dr'\,W(r,r')\oper{L}_{r'}\Bigl(Q(\epsilon^{-1}r')a(z;r')\Bigr)\;.
\]
Since $a$ is a slowly varying function of $r'$, its radial
derivatives are of order unity. The dominant term in $\oper{L}_{r'}Q$
arises from $d^2Q/dr'^2=O(\epsilon^{-2})$. Since the
magnitude of the integral is controlled by the singularity in
$W(r,r')$, Eq.~(\ref{Wsing}), we have
\[
\oper{K}(Qa)\simeq\int_0^\infty \ln|r-r'|\frac{d^2Q}{dr'^2}\,dr'\;.
\]
It may seem that the order of magnitude of this integral is
$\epsilon^{-2}\Delta r\ln|\Delta r|\simeq\epsilon^{-1}\ln\epsilon$
because the main contribution to the integral arises from a
neigbourhood of $r=r'$ whose width is $\Delta r\simeq\epsilon$.

However, this estimate is wrong because $d^2Q/dr'^2$
changes rapidly and even changes sign within the
$\epsilon$-neighbourhood of $r-r'$. The correct calculation is as
follows. We introduce $\delta\to0$ and consider
\[
\int_{0}^\infty \ln|r-r'|\frac{d^2Q}{dr'^2}\,dr'
\simeq\int_0^{r-\delta}  \ln|r-r'|\frac{d^2Q}{dr'^2}\,dr'
+     \int_{r+\delta}^\infty\ln|r-r'|\frac{d^2Q}{dr'^2}\,dr'\;,
\]
so that now we can integrate by parts to obtain
\[
\oper{K}(Qa)
=\left[\frac{dQ}{dr'}\ln|r-r'|\right]_{0}^{r-\delta}+
 \left[\frac{dQ}{dr'}\ln|r-r'|\right]_{r+\delta}^\infty
-{\rm v.p.}\int_0^\infty\frac{dQ}{dr'}\frac{1}{r-r'}\,dr'\;.
 \]
Since the $\ln|r-r'|$ is symmetric with respect to
$r,r'$ and $dQ/dr=0$ for
$r=0$ and $\infty$, the logarithmic terms cancel and
\[
\oper{K}(Qa)
=-{\rm
v.p.}\int_0^\infty\frac{dQ}{dr'}\frac{1}{r-r'}\,dr'\simeq\epsilon^{-1}\;.
\]

\section*{Appendix C. The adjoint problem}

\setcounter{equation}{0}
\renewcommand\theequation{C\arabic{equation}}

Consider the homogeneous counterpart of the boundary
value problem (\ref{Ezf}) and (\ref{Bzf}), i.e.,
\begin{equation}
\oper{L}_z \vect{X} - \gamma_0 \vect{X} = 0\;,
\label{Emhom}
\end{equation}
\begin{equation}
\oper{P}_{z,0}\vect{X}=0\;, \quad \oper{P}_{z,1} \vect{X} =0\;.
\label{Bmhom}
\end{equation}
This boundary value problem coincides with the zeroth-order problem
(\ref{Eqzero}) and (\ref{BCzero}).
Our aim here is to formulate the adjoint problem and to discuss
its properties.

We introduce the eigenvectors of the original and adjoint problems,
respectively:
\[
\vect{X} = {{b(z;r)}\choose {a(z;r)}}, \quad
\vect{X}^{\dag} = {{b^{\dag}(z;r)}\choose {a^{\dag}(z;r)}}\;.
\]
By the definition of the adjoint operator, we have
\begin{eqnarray*}
\langle\oper{L}_z\vect{X},\vect{X}^{\dag}\rangle &=&
\int_0^1 \left[(-DGa'+b'')b^{\dag}+(\alpha b+a'')a^{\dag}\right]\,dz\\
&=&\left.\left(-DGab+b'b-b{b^{\dag}}'\right)\right|^1_0
        + \left.\left(a'a^{\dag} - a{a^{\dag}}'\right)\right|^1_0+\\
&&\quad\mbox{}+\int_0^1 \left[b \left({b^{\dag}}''+ \alpha a^{\dag}\right) + a
\left(DG{b^{\dag}}'+{a^{\dag}}''\right)\right]\, dz\\
&=&\langle\vect{X},\oper{L}_z^{\dag} \vect{X}^{\dag}\rangle \;.
\end{eqnarray*}

Therefore, the problem adjoint to (\ref{Eqzero}), (\ref{BCzero}) is
\begin{eqnarray*}
\gamma_0^{\dag}(r)b^{\dag}&=&{b^{\dag}}''+\alpha a^{\dag}\;,\\
\gamma_0^{\dag}(r)a^{\dag}&=& DG{b^{\dag}}'+{a^{\dag}}''\;,
\end{eqnarray*}
\begin{equation}
a^{\dag}(0)=0\;, \quad
{b^{\dag}}'(0)=0\;,\quad b^{\dag}(1)=0\;, \quad {a^{\dag}}'(1)=0\;.
                                                \label{BCadj}
\end{equation}
According to Fredholm's theory, the
adjoint problem has the same spectrum as the problem
(\ref{Eqzero}) and (\ref{BCzero}), i.e., ${\gamma_0}^{\dag}(r) = \gamma_0$.

\section*{References}

\begin{description}

\item
Belyanin, M.P., Sokoloff D.D., and Shukurov, A.M.\ (1994)
Asymptotic steady-state solutions to the nonlinear
hydromagnetic dynamo equations, {\it Russ.\ J.\ Math.\ Phys.,}
{\bf 2}, 149--174.

\item
Kolmogorov, A.N., and Fomin, S.V.\ (1957) {\it Elements of the Theory
of Functions and Functional Analysis,}
Graylock Press, Rochester, N.Y.

\item
Moss, D., Shukurov, A., and Sokoloff, D.\ (1998) Boundary effects and
propagating magnetic fronts in disc dynamos, {\it Geophys.\
Astrophys.\ Fluid Dyn.}, {\bf 89}, 285--308.

\item
Poezd, A., Shukurov, A.,  and Sokoloff, D.\ (1993) Global magnetic
         patterns in the Milky Way and the Andromeda nebula. {\it Mon.\
         Not.\ Roy.\ Astron.\ Soc.\/}, {\bf 264}, 285--297.

\item
Ruzmaikin, A.A., Shukurov, A.M., and Sokoloff, D.D.\ (1988) {\it
Magnetic Fields of Galaxies}, Kluwer, Dordrecht.

\item
Soward, A.M.\ (1978) A thin disc model of the Galactic dynamo, {\it
Astron.\ Nachr.}, {\bf 299}, 25--33.

\item
Soward, A.M.\ (1992a) This disc $\alpha\omega$-dynamo models. I. Long
length scale modes,  {\it Geophys.\
Astrophys.\ Fluid Dyn.}, {\bf 64}, 163--199.

\item
Soward, A.M.\ (1992b) This disc $\alpha\omega$-dynamo models I. Long
length scale modes,  {\it Geophys.\
Astrophys.\ Fluid Dyn.}, {\bf 64}, 201--225.

\item
Soward, A.M.\ (2000) Thin aspect ratio $\alpha\Omega$-dynamos in
galactic discs and stellar shells, in: {\it Advances in Nonlinear
Dynamos\/} (Ed.\ M.~N\'u\~nez and A.~Ferriz Mas), {\it The Fluid
 Mechanics of Astrophysics and Geophysics}, Gordon and Breach, New
 York (to appear).

\item
Sneddon, I.N.\ (1951) {\it Fourier Transforms}, McGraw--Hill, N.Y.

\item
Willis, A., Shukurov, A., Sokoloff, D.D., and Soward, A.M.\ (2000)
Non-local effects in the mean-field disc dynamo. II. (in preparation).

 \item
Zeldovich, Ya.B., Ruzmaikin, A.A., and Sokoloff, D.D.\ (1983)
 {\it Magnetic Fields in Astrophysics,} Gordon and Breach, New York.

\end{description}

\end{document}